\documentclass[final,5p,times,twocolumn]{elsarticle}

\usepackage{color}
\usepackage{graphics}
\usepackage{amsmath}

\begin{document}

\begin{frontmatter}

\title {Vacuum torsion and regular accelerating Universe without dark matter}%

\author{A.~V.~Minkevich}

\ead{minkav@bsu.by, awm@matman.uwm.edu.pl}

\address{Department of Theoretical Physics and Astrophysics, Belarussian State
University, Minsk, Belarus} \address{Department of Physics and Computer
Methods, Warmia and Mazury University in Olsztyn, Poland}

%\date{\today}

\begin{abstract}
The simplest gauge gravitation theory in Riemann-Cartan space-time leading
to the solution of the problem of cosmological singularity and dark energy
problem is investigated with purpose to solve the dark matter problem. It
is shown that the interaction of the vacuum torsion with proper angular
momentums of gravitating objects can lead to appearance at astrophysical
scale (galaxies, galactic clusters) additional force of gravitational
attraction, which in the frame of standard theory is associated with dark
matter.
\end{abstract}


\end{frontmatter}

\section {Introduction}

The application of the local gauge invariance principle, which is one of the most
important principles of modern theory of fundamental physical interactions, generally
speaking leads to generalization of metric gravitation theory (MGT). MGT can be
introduced in the frame of gauge approach by considering the 4-translations group as
gauge group \cite{1, 2}. By including the Lorentz group into gauge group
corresponding to gravitational interaction we obtain gravitation theory in
Riemann-Cartan space-time (GTRC) \cite {3, 4, 5} \footnote{Now there are many works
dedicated to investigations of GTRC, which is known in literature as Poincar\'e gauge
theory of gravity (see e.g. \cite{5a, 5b, 5c, 5d, 17, 12} and Refs herein).}. It should be
noted that GTRC but not MGT corresponds to supergravity theory, because the
supergravity gauge group includes the Lorentz group. If GTRC is natural generalization
of MGT, the question arose about the possibility of the solution of principal problems
of general relativity theory (GR) on the base of GTRC, at first of all the problem of the
beginning of the Universe in time and problem of dark components of the Universe -
dark energy and dark matter. In the frame of GTRC based on sufficiently general
expression of gravitational Lagrangian ${\cal L}_{\rm g}$ including both a scalar
curvature and terms quadratic in the curvature and torsion tensors with indefinite
parameters it was shown that by certain restrictions on these parameters GTRC allows
to solve the problem of the beginning of the Universe in time (problem of cosmological
singularity) and to explain accelerating cosmological expansion at present epoch
without using the notion of dark energy. It is because GTRC leads to the change of
gravitational interaction by certain conditions in comparison with GR: gravitational
interaction has repulsive character at extreme conditions (extremely high energy
densities and pressures) where limiting energy (mass) density appears \cite{6} and in
situation when energy density is very small and the vacuum gravitational repulsion
effect is essential \cite{7a, 7}. These physical results were obtained on the base of
investigations of isotropic cosmology built in the frame of GTRC (see e.g. \cite{6, 7a, 7,
8, 9, 11}). The physical significance of space-time torsion in dynamics of gravitating
systems was found out. In particular, the principal role of torsion is connected with the
fact that physical space-time in the vacuum has the structure of Riemann-Cartan
continuum with de Sitter metric (but not Minkowski space-time) \cite{7}.

The present letter is devoted to investigation of dark matter problem on the
base of GTRC: is it possible to explain phenomena associated in the frame of
standard theory with dark matter as a result of the change of gravitational
interaction at astrophysical scale in comparison with GR? Our consideration
will be realized in the frame of so-called minimum GTRC - the simplest GTRC,
which allows to solve the problem of cosmological singularity and dark
energy problem.

\section {Minimum gauge gravitation theory in Riemann-Cartan space-time and vacuum torsion}

The minimum GTRC \cite {11, 13} was introduced as a result of investigations
of GTRC based on general expression of gravitational Lagrangian ${\cal
L}_{\rm g}$ including both a scalar curvature $F$, six invariants quadratic
in the curvature tensor $F^{ik}{}_{\mu\nu}$ and three invariants quadratic
in the torsion tensor $S^i{}_{\mu\nu}$ with indefinite parameters
\footnote{The definitions and notations of our previous papers (see e.g.
\cite{7a, 11, 13}) are used below. With the purpose to make quantitative
estimations the light velocity $c$ is conserved in formulas.}. Restrictions
on indefinite parameters of ${\cal L}_{\rm g}$ were obtained, by which the
physical consequences of GTRC are the most satisfactory. As a result the
following expression of ${\cal L}_{\rm g}$ of minimum GTRC was found:
\begin{eqnarray} \label{1}
{\cal L}_{\rm g}=f_0\,
F+ f_5\: F^{\mu\nu} F_{\nu\mu} +  f_6\:F^2
\nonumber \\
        +S^{\alpha\mu\nu} (a_1\:S_{\alpha\mu\nu}+a_2\:
        S_{\nu\mu\alpha})
    +a_3\:S^\alpha{}_{\mu\alpha}S_\beta{}^{\mu\beta},
\end{eqnarray}
where $f_0=\frac{c^{4}}{16\pi G}$ ($G$ is Newton's gravitational constant)
and parameters $f_5$, $f_6$ and $a_k$ ($k=1,2,3$) are expressed in terms of
parameters of isotropic cosmology: $b=a_2-a_1$,
$\alpha=\frac{f_5+3f_6}{3f_0^2}>0$ and $\omega=\frac{f_5}{f_5+3f_6}$
\cite{6,13}:
\begin{eqnarray}\label{2.11}
a_1=b,\qquad a_2=2b,\qquad a_3=-\frac{4}{3}b,
\nonumber \\
 f_5=3f_0^2 \alpha \omega,\qquad f_6=f_0^2 \alpha (1-\omega).
\end{eqnarray}
The correspondence principle with GR will be fulfilled if parameters $b$ and
$\omega$ satisfy the following restrictions: $0<1- \frac{b} {f_0}\ll 1$,
$0<\omega\ll 1$ and the value of $\alpha^{-1}$ corresponds to some high
energy density (see below).

Gravitational equations of minimum GTRC have the following form:
\begin{eqnarray}\label{3.1}
\nabla_{\nu}U_{i}{}^{\mu\nu}+2S^k{}_{i\nu}U_k{}^{\mu\nu}+
2(f_0+2f_6\:F)F^{\mu}{}_i
\nonumber\\
+2f_5(F_{ki}F^{\mu k}+F^{\mu}{}_{kim}F^{mk})-
h_{i}{}^{\mu}(f_0 F+ f_5 F^{\mu\nu} F_{\nu\mu} + f_6\:F^2
\nonumber\\
+S^{\alpha\mu\nu}\left(a_1\:S_{\alpha\mu\nu}+a_2\: S_{\nu\mu\alpha}\right)
+a_3\:S^\alpha{}_{\mu\alpha}S_\beta{}^{\mu\beta}) =-T_{i}{}^{\mu},
\end{eqnarray}
\begin{eqnarray}\label{3.2}
4\nabla_{\nu}[(f_0/2+f_6\:F)h_{[i}{}^{\nu}h_{k]}{}^{\mu}+
\nonumber\\
+f_5\:F^{[\mu}{}_{[k}h_{i]}{}^{\nu]}]+U_{[ik]}{}^{\mu}=-J_{[ik]}{}^{\mu},
\end{eqnarray}
where
$U_{i}{}^{\mu\nu}=2(a_1\:S_{i}{}^{\mu\nu}-a_2\:S^{[\mu\nu]}{}_{i}-a_3\:S_{\alpha}{}^{\alpha
[\mu }h_{i}{}^{\nu]})$, $\nabla_{\nu}$ denotes the covariant operator having
the structure of the covariant derivative defined in the case of tensor
holonomic indices by means of Christoffel coefficients and in the case of
tetrad tensor indices by means of anholonomic Lorentz connection.

As it was shown in \cite {13}, if $0<\omega\ll 1$ in the case of spinless
matter with energy densities, which are much less than limiting energy
density, equations (3)-(4) lead to equations for metric in the form of
Einstein gravitational equations with effective cosmological constant
\begin{equation}\label{3.14}
G^{\mu}{}_{\lambda}=-\frac{1}{2b}
 \left[T_{\lambda}{}^{\mu} + \delta^{\mu}_{\lambda}
\frac{(1-\frac{b}{f_0})^2}{12\alpha} \right],
\end{equation}
where $G^{\mu}{}_{\lambda}$ is Einstein tensor. The influence of torsion
appears in eq. (5) via formation of effective cosmological constant and the
change of gravitational constant. We see that the correspondence principle
with GR will be fulfilled if parameters $b$ satisfies the condition $0<1-
\frac{b} {f_0}\ll 1$ and the value of $\alpha^{-1}$ corresponds to some high
energy density, by which effective cosmological constant ensures observed
accelerating cosmological expansion.

As it follows from equations of isotropic cosmology the effective
cosmological constant in equations (5) has the vacuum origin. Let us
illustrate this fact. Any homogeneous isotropic model (HIM) in
Riemann-Cartan space-time is described by three functions of time: the scale
factor of Robertson-Walker metric $R(t)$ and two torsion functions
$S_{1}(t)$ and $S_{2}(t)$. Cosmological equations generalizing Friedmann
cosmological equations of GR take the form \cite{6}
\begin{eqnarray}\label{2.2}%\fl
    \frac{k}{R^2} + (H-2S_1)^2 -S_2^2= \nonumber\\
    \frac{1}{{6f_0 Z}}
        \left[
            {\rho c^2  -6 b S_2^2
            + \frac{\alpha }{4} \left( {\rho c^2  - 3p - 12bS_2^2 } \right)^2 }
        \right],
\end{eqnarray}
\begin{eqnarray}\label{2.3}%\fl
    \dot{H}-2\dot{S}_1 +H (H-2S_1)= \nonumber\\
    -\frac{1} {{12f_0 Z}}
        \left[
            \rho c^2  + 3p - \frac{\alpha } {2} \left( {\rho c^2 - 3p - 12bS_2^2 } \right)^2
        \right],
\end{eqnarray}
where $H=\dot{R}/R $ is the Hubble parameter (a dot denotes the
differentiation with respect to $x^0= c t$), $k=+1,0,-1$ for closed, flat
and open models respectively, $\rho$ is mass density, $p$ is pressure and
$Z=1+\alpha\left( \rho c^2 - 3p - 12b S_2^2\right)$. The torsion functions
$S_1$ and $S_2$ are
\begin{eqnarray}\label{2.4}%\fl
    S_1  = -\frac{\alpha }{4Z} [\dot \rho c^2
    - 3 \dot p + 12f_0 \omega H S_2^2
    -12( {2b - \omega f_0 } ) S_2 \dot S_2],
\end{eqnarray}
\begin{eqnarray}\label{2.5}
 S_{2}^{2}  = \frac{\rho c^2 - 3p}{12b} + \frac
{1-(b/2f_0) (1 +  \sqrt{X})} {12b \alpha (1- \omega/4)},
\end{eqnarray}
where
\begin{equation}\label{2.6}
X=1+ \omega (f_0^2/b^2) [1- (b/f_0) - 2(1- \omega /4) \alpha ( \rho c^2+ 3p)]\ge
0.
\end{equation}

The dynamics of HIM depends essentially on torsion functions. The presence
of $\sqrt{X}$ in (10) leads to appearance of limiting energy density of
order $(\omega\alpha)^{-1}$ and secures regular behaviour of HIM at extreme
conditions, and the presence of constant term in $S_{2}^{2}$ - vacuum
torsion - induces the cosmological constant at asymptotics. By taking into
account restrictions for parameters $\omega$ and $b$ the expression (9) for
$S_{2}^{2}$ at asymptotics takes the form:
\begin{equation}\label{2.8}
S_2^2  = \frac {1} {12b} \left[\rho c^2  - 3p + \frac {1  - b/f_0} {\alpha}\right],
\end{equation}
and as a result cosmological equations (6)-(7) at asymptotics are:
\begin{equation}\label{2.9}
    \frac{k}{R^2 } + H^2  = \frac{1}{6b }\left[\rho c^2 + \frac{1}{4\alpha} \left(1 - \frac{b}{f_0}\right)^2
     \right],
\end{equation}
\begin{equation}\label{2.10}
    \dot H + H^2  =  - \frac{1} {{12b }}\left[ (\rho c^2 + 3p) - \frac{1}{2\alpha}
    \left(1 - \frac{b}{f_0}\right)^2 \right].
\end{equation}
Unlike standard $\Lambda CDM$-model cosmological constant appears in
(12)-(13) as a result of solution of gravitational equations for HIM that
leads to the change of gravitational interaction when energy density is
small and comparable with cosmological constant - the vacuum gravitational
repulsion effect. This effect appears at cosmological scale, and it is
negligibly small at astrophysical scale and locally (Solar system).

The influence of gravitating vacuum in GTRC can be manifested not only via
accelerating cosmological expansion at present epoch. The fact of the matter
is that the vacuum torsion function $S_{2}^{2(vac)}$, which according to
(11) is
\begin{equation}\label{2.8}
S_{2}^{2(vac)}  =  \frac {1  - b/f_0} {12b \alpha},
\end{equation}
can be essential quantitatively at newtonian approximation. The vacuum torsion
function $|S_1|$ and the vacuum value of $H$ are negligibly small in comparison with
$|S_{2}^{(vac)}|$. Owing to this the curvature tensor \cite{7a} has the following
vacuum components, which can be important at newtonian approximation:
\begin{equation}
F^{12}{}_{12} = F^{13}{}_{13} = F^{23}{}_{23} = - S_{2}^{2(vac)}.
\end{equation}
It is easy to show by using \cite{13} that according to eq. (4) corrections
of the torsion components $S_{\alpha\mu\nu}$ ($\alpha,\mu,\nu =1,2,3$,
${\alpha} \neq {\mu}$, ${\alpha} \neq {\nu}$) and consequently curvature
components (15) at asymptotics in the case of matter with spin are not
essential at newtonian approximation. It should be noted that though the
formula (14) is obtained in cosmological concomitant system of reference,
formulas (14) and (15) are valid by transition to other systems of reference
because $S_2^2=-\frac{1}{6} S_{\alpha\mu\nu}S^{\alpha\mu\nu}$
($\alpha,\mu,\nu =1,2,3$; ${\alpha} \neq {\mu}$, ${\alpha} \neq {\nu}$).

\section {Vacuum torsion and gravitational interaction at astrophysical scale}

Lets consider the interaction of proper angular momentums of astrophysical
objects (stars in galaxies, galaxies in galactic clusters) with vacuum
torsion by using equations of motion of particle with momentum in
Riemann-Cartan space-time \cite{14} generalizing Papapetrou's equations for
rotating particle in GR \cite{15} \footnote {In \cite{14} the curvature
tensor was defined with opposite sign and signature (+2) was used.}. In the
case of rotating particle with angular velocity tensor $\Omega_{ik}$
corresponding equations of motion, by conserving terms, which are essential
at non-relativistic approximation, are:
\begin{equation}\label{2.8}
\frac {D P_i} {d \tau} = \frac{1}{2} I \Omega_{mn} F^{mn}{}_{il} v^{l} \qquad (i,l,m,n=1,2,3),
\end{equation}
where $\frac{D}{d \tau}$ denotes riemannian absolute derivative with respect
to proper time $\tau$, $P_i$ is generalized momentum, $I$ is inertia
momentum  and $v^{l}$ is velocity of particle. In non-relativistic
approximation $P_i= m v_{i}$ ($m$ is particle mass) and $\Omega_{mn}=const$.

We will consider the circular motion of rotating particle in spherically
symmetric gravitational field created by mass $M$ in newtonian
approximation. By taking into account that $g_{00}= 1+\frac{2\phi}{c^2}$
($\phi$ is newtonian potential), components of angular velocity $\Omega_{i}=
\epsilon_{ikl} \Omega^{kl}$  and relation (15) we obtain in the case of
motion in plane $XOY$ (centrum of mass $M$ is in origin of coordinates,
vector of orbital angular momentum is directed along the axe $OZ$) equation
of motion in usual form $m \frac {d\mathbf{v}} {dt} = \mathbf{F}$ with the
following expression of the force vector:
\begin{equation}\label{2.8}
\mathbf{F}=-m\frac{d\phi}{d\mathbf{r}} + I \Omega_3 S_{2}^{2(vac)} v \frac{\mathbf{r}}{r}.
\end{equation}
If $\Omega_3 < 0$ the force (17) includes besides Newtonian term additional force of attraction:
\begin{equation}\label{2.8}
F=G\frac{mM}{r^2} + I \Omega S_{2}^{2(vac)} v ,
\end{equation}
where $\Omega= |\Omega_3|$. By taking into account that the force (17) is
centripetal force we obtain the following dependence of velocity on distance
from centrum and parameters of particle and gravitational field:
\begin{equation}\label{2.8}
v= \frac{I}{2m} \Omega S_{2}^{2(vac)} r + \left[(\frac{I}{2m} \Omega S_{2}^{2(vac)} r)^2 + \frac{GM}{r}\right]^{\frac{1}{2}}.
\end{equation}
We see that interaction of proper angular momentum with vacuum torsion leads
to terms growing with distance in expression of velocity (19). This allows
to explain the behaviour of rotational curves in galaxies. By given
parameters of particle and gravitational field the velocity (19) depends on
value of parameter $x=1- \frac{b} {f_0}$. By taking into account that
average mass density in the Universe at present epoch $\rho_1
=\frac{x^2}{4c^2 \alpha}$ is of order $10^{-26}
\frac{\text{kg}}{\text{m}^3}$, we obtain that at the first approximation
\begin{equation}\label{2.8}
S_{2}^{2(vac)}=\frac{16 \pi G}{3c^2 x} \rho_1 \sim {\frac{10^{-52}}{x}} (\text{m}^{-2}).
\end{equation}
Lets demonstrate the behaviour of rotational curves numerically by applying
obtained relation (19) for hypothetical galaxy similar to Andromeda by
choosing particle parameters as for star similar to Solar: $I/m\sim10^{18}$
$\text{m}^2$, $\Omega \sim 0.5\cdot10^{-6}$ $\text{s}^{-1}$; the mass $M$ is
taken $M = 2\cdot10^{41}$ $\text{kg}$; the parameter $x =10^{-25}$ and
consequently $S_{2}^{2(vac)}= 10^{-27} (\text{m}^{-2})$ that corresponds to
high energy density scale $\alpha^{-1} = 10^{7} \rho_{nucl}c^2$
($\rho_{nucl}$ is nuclear mass density). As numerical analysis shows at
distances $r < 9$ kpc ($1 $\text{kpc}$=0,3086\cdot10^{20} \text{m}$)
Newtonian term in (19) plays the definitive role, by growth of $r$ from 9
kpc to 25 kpc the velocity $v$ according to Newtonian law decreases from
$219\cdot10^{3}$ km/s to $132\cdot10^{3}$ km/s, but according to (19) the
velocity $v$ changes only from $256\cdot10^{3}$ km/s to $259\cdot10^{3}$
km/s. By further increase of $r$ essential growth of velocity $v$ takes
place according to (19); this effect can be observed in galactic clusters,
where we deal with vast space scale of order $10 \text{Mpc}$ and more. At
the same time this effect can explain why in the frame of standard theory
the motion of Dwarf galaxies in galactic clusters is subjected to dark
matter, but dark matter does not have an effect on motion in the interior of
Dwarf galaxies. If we take into account that linear dimensions of Solar
system compose small part of parsec, effects discussed above at galactic
scale become negligibly small in Solar system.

Investigation of influence of vacuum torsion on propagation of
electromagnetic waves is interesting in connection with observations of
gravitational lensing. Such influence is possible if gravitational
interaction with electromagnetic field in the frame of GTRC is given by
using the minimal coupling (replacing in Lagrangian of electromagnetic field
written in Minkowski space-time of partial derivatives via covariant
derivatives defined with help of total connection). Because the minimal
coupling leads to violation of gauge invariance in the case of
electromagnetic field by conservation of electric charge, gravitational
interaction in the frame of GTRC generally is specified similar to GR by
using Christoffel coefficients as connection. However, the minimal coupling
rather leads in this case to partially gauge theory in terms of \cite {16}
that is in agreement with interaction hierarchy and provides an interesting
results \cite{17}.

\section {Conclusion}

The physical vacuum in the frame of GTRC is gravitating system possessing the
curvature and torsion and having the structure of Riemann-Cartan continuum with de
Sitter metric. This conclusion was obtained in the frame of theory based on general
expression of gravitational Lagrangian without using any restrictions on indefinite
parameters of ${\cal L}_{\rm g}$ \cite{7}. Physical consequences of this result at first
were obtained in the frame of isotropic cosmology, where the vacuum gravitational
repulsion effect and connected with them accelerating cosmological expansion at
present epoch were described. However, investigation of possible role of gravitating
vacuum at astrophysical scale is difficult task through complexity of gravitational
equations of GTRC. The situation is simplifying, when minimum GTRC was determined.
It is because this theory leads to Einstein gravitation equations for metric, which are
valid for spinless gravitating systems at wide range of energy density - from extremely
high energy densities defined by $\alpha^{-1}$ to energy densities, which are several
order greater than average energy density in the Universe at present epoch. This
means that Newton's law of gravitational attraction is valid for spinless matter at
astrophysical scale in non-relativistic approximation. Corrections connected with spin
effects of gravitating matter in the frame of GTRC locally are very small \cite{18}. The
possible principal role of gravitating vacuum in astrophysics was determined in this
letter. As it is shown rotational curves in galaxies and gravitational phenomena in
galactic clusters, for explanation of which the notion of dark matter was introduced in
the frame of GR, can be explained in the frame of GTRC as a result of interaction of
proper angular momentums of stars and galaxies with torsion of gravitating vacuum
\footnote {Because discussed phenomena quantitatively are essential at newtonian
approximation, the search similar to \cite{18} of possible experiments for observations
of interaction of proper angular momentums of gyroscopes (similar to Pioneer 10 and
Pioneer 11) with vacuum torsion takes on new significance.}. We see that physical
phenomena associated in the frame of GR with dark matter as well as dark energy have
in the frame of GTRC the vacuum origin. This differs our results from that proposed in
the frame of MOND \cite{19} and other alternative theories of gravity, where
explanation of discussed phenomena is given by means of modified newtonian
gravitational potential. Of course the research of rotational curves for real galaxies and
study of motions in galactic clusters suppose considerable information relative to
galaxies and galactic clusters. As a result values of parameters $b$ and $\alpha$ can be
defined, and only parameter $\omega$, which determines the value of limiting energy
density, limiting temperature and depending on them  the state and physical processes
in the Universe at the beginning of cosmological expansion remains undefined
\footnote {In the frame of our classical theory limiting energy density has to be less
than the Planckian one. The existence of limiting energy density ensures the regular
behaviour of all HIM, including inflationary cosmological models \cite{8,9}.}.

\end{document}